\documentclass[%
 reprint,
 superscriptaddress,
 amsmath,amssymb,
 aps,
 prb,
]{revtex4-2}

\setcitestyle{number}
\usepackage{graphicx}
\usepackage{amsmath,amsfonts,amssymb}
\usepackage[dvipsnames]{xcolor}
\usepackage{dcolumn}
\usepackage{bm}
\usepackage{hyperref}
\hypersetup{
    colorlinks=true,
    linkcolor=blue,
    filecolor=magenta,      
    urlcolor=blue,
    citecolor=blue
    }

\usepackage{blindtext}
\usepackage{multirow}

\begin{document}

\title{BaZrS$_\text{3}$ Lights Up: The Interplay of Electrons, Photons, and Phonons in Strongly Luminescent Single Crystals}

\author{Rasmus Svejstrup Nielsen}
\email[]{Electronic mail: rasmus.nielsen@empa.ch}
\affiliation{Nanomaterials Spectroscopy and Imaging, Transport at Nanoscale Interfaces Laboratory, Swiss Federal Laboratories for Material Science and Technology (EMPA), Ueberlandstrasse 129, 8600 Duebendorf, Switzerland}

\author{Ángel Labordet Álvarez}
\affiliation{Nanomaterials Spectroscopy and Imaging, Transport at Nanoscale Interfaces Laboratory, Swiss Federal Laboratories for Material Science and Technology (EMPA), Ueberlandstrasse 129, 8600 Duebendorf, Switzerland}
\affiliation{Department of Physics, University of Basel, 4056 Basel, Switzerland}
\affiliation{Swiss Nanoscience Institute, University of Basel, 4056 Basel, Switzerland}

\author{Yvonne Tomm}
\affiliation{Department of Structure and Dynamics of Energy Materials, Helmholtz-Zentrum Berlin für Materialien und Energie, Hahn-Meitner-Platz 1, 14109 Berlin, Germany}

\author{Galina Gurieva}
\affiliation{Department of Structure and Dynamics of Energy Materials, Helmholtz-Zentrum Berlin für Materialien und Energie, Hahn-Meitner-Platz 1, 14109 Berlin, Germany}

\author{Andres Ortega-Guerrero}
\affiliation{Nanotech@surfaces Laboratory, Swiss Federal Laboratories for Material Science and Technology (EMPA), Ueberlandstrasse 129, 8600 Duebendorf, Switzerland}

\author{Joachim Breternitz}
\affiliation{Department of Structure and Dynamics of Energy Materials, Helmholtz-Zentrum Berlin für Materialien und Energie, Hahn-Meitner-Platz 1, 14109 Berlin, Germany}
\affiliation{FH Münster, Department of Chemical Engineering, Stegerwaldstr. 39, 48565 Steinfurt, Germany}

\author{Lorenzo Bastonero}
\affiliation{U Bremen Excellence Chair, Bremen Center for Computational Materials Science, and MAPEX Center for Materials and Processes, University of Bremen, D-28359 Bremen, Germany}

\author{Nicola Marzari}
\affiliation{U Bremen Excellence Chair, Bremen Center for Computational Materials Science, and MAPEX Center for Materials and Processes, University of Bremen, D-28359 Bremen, Germany}
\affiliation{PSI Center for Scientific Computing,
Theory, and Data, and National Centre for Computational Design and Discovery of Novel Materials (MARVEL), 5232 Villigen PSI, Switzerland}
\affiliation{Theory and Simulation of Materials (THEOS), and National Centre for Computational Design and Discovery of Novel Materials (MARVEL), \'Ecole Polytechnique F\'ed\'erale de Lausanne (EPFL), CH-1015 Lausanne, Switzerland}

\author{Carlo Antonio Pignedoli}
\affiliation{Nanotech@surfaces Laboratory, Swiss Federal Laboratories for Material Science and Technology (EMPA), Ueberlandstrasse 129, 8600 Duebendorf, Switzerland}

\author{Susan Schorr}
\affiliation{Department of Structure and Dynamics of Energy Materials, Helmholtz-Zentrum Berlin für Materialien und Energie, Hahn-Meitner-Platz 1, 14109 Berlin, Germany}
\affiliation{Institute of Geological Sciences, Freie Universität Berlin, Malteserstr. 74–100, 12249 Berlin, Germany}

\author{Mirjana Dimitrievska}
\email[]{Electronic mail: mirjana.dimitrievska@empa.ch}
\affiliation{Nanomaterials Spectroscopy and Imaging, Transport at Nanoscale Interfaces Laboratory, Swiss Federal Laboratories for Material Science and Technology (EMPA), Ueberlandstrasse 129, 8600 Duebendorf, Switzerland}

\begin{abstract}\vspace{0.2cm} 

Chalcogenide perovskites have emerged as a promising class of materials for the next generation of optoelectronic applications, with BaZrS$_\text{3}$ attracting significant attention due to its wide bandgap, earth-abundant composition, and thermal and chemical stability. However, previous studies have consistently reported weak and ambiguous photoluminescence (PL), regardless of synthesis method, raising questions about the intrinsic optoelectronic quality of this compound. In this work, we demonstrate strong, band-to-band-dominated PL at room temperature in high-quality BaZrS$_\text{3}$ single crystals, with a PL quantum yield of $\sim$0.005\%. Despite the narrow, single-component PL emission band, time-resolved PL measurements reveal a carrier lifetime of $1.0\pm0.2$ ns. To understand the origin of the strong PL and short carrier lifetime, we perform multiwavelength excitation and polarization-dependent Raman measurements, supported by first-principles lattice dynamics calculations. We identify all 23 theoretically predicted Raman-active modes and their symmetries, providing a comprehensive reference for future studies. Our results indicate that phonon-assisted carrier decay and nontrivial electron-phonon interactions contribute to the short carrier lifetimes, as evidenced by Raman spectroscopy and DFT calculations. Further studies on compositional variations or partial cation/anion substitutions could mitigate electron-phonon coupling and enhance carrier lifetimes. By establishing a detailed reference for the intrinsic vibrational and optoelectronic properties of BaZrS$_\text{3}$, this work paves the way for further advancements in chalcogenide perovskites for energy and optoelectronic technologies.

\end{abstract}

\maketitle

\section{Introduction}

The emergence of lead-halide perovskites has transformed the field of optoelectronics \cite{wang2021a, liu2021a, ren2022a, lei2021a, cheng2022a}, driven by their exceptional optoelectronic properties, including high absorption coefficients, long carrier diffusion lengths, and defect tolerance \cite{stranks2013a, dequilettes2016a, de2014a}. However, their practical implementation faces significant challenges due to their thermal and chemical instability \cite{duan2023a, conings2015a, leijtens2015a}, as well as concerns over the toxicity of lead \cite{babayigit2016a, serrano2015a}. In response, chalcogenide perovskites have gained attention as a promising alternative, offering superior stability and environmentally benign compositions \cite{niu2018a}, with bandgaps within the visible region that are relevant for energy conversion and light technologies \cite{agarwal2025a, han2024a, yaghoubi2024a}. Among these materials, BaZrS$_\text{3}$ has been the most widely studied, owing to its wide direct bandgap and its ability to be synthesized at comparatively lower temperatures than other chalcogenide perovskites \cite{pradhan2023a, raj2023a, crovetto2019a, comparotto2025a, vincent2023a}.

BaZrS$_\text{3}$ exhibits a high absorption coefficient ($>10^5$ cm$^\text{-1}$ in the visible region) \cite{cheng2022a, wei2019a}, stability under ambient conditions \cite{niu2018a, xu2022a}, and is composed of earth-abundant elements \cite{vesborg2012a}, making it a promising candidate for a range of optoelectronic applications \cite{agarwal2025a}. However, the optoelectronic properties remain debated in the literature, with the few existing studies reporting weak and ambiguous photoluminescence (PL). \cite{ye2024a, niu2017a, wei2019a, surendran2021a, comparotto2020a, ye2022a, pradhan2023a, yang2022a}. The reported PL spectra vary significantly, showing inconsistencies in peak positions, asymmetric emission bands, multi-component features, and, in some cases, a complete absence of room-temperature PL. Since strong, well-defined PL is a key indicator of high optoelectronic quality, these discrepancies raise fundamental questions about intrinsic limitations to the optoelectronic potential of BaZrS$_\text{3}$. As the origins of these variations remain unclear -- they may result from quenched-in defects introduced during synthesis or strong electron-phonon coupling facilitating non-radiative recombination -- a detailed study of the optoelectronic and vibrational properties of high-crystal quality BaZrS$_\text{3}$ is needed to assess the role of both defects and phonons. 

In a recent study, Ye et al. compared the vibrational properties of the lead-halide perovskite CsPbBr$_\text{3}$ and BaZrS$_\text{3}$ by investigating temperature-dependent PL and Raman spectra \cite{ye2024a}. They concluded that electron-phonon coupling in BaZrS$_\text{3}$ is significantly stronger, likely due to greater phonon anharmonicity. However, their PL measurements revealed multiple spectral components and no detectable signal at room temperature, suggesting poor optoelectronic quality in their sample. Yetkin et al. combined theoretical and experimental Raman studies on BaZrS$_\text{3}$-BaHfS$_\text{3}$ solid solutions, showing that non-resonant Raman spectra varied only slightly with composition, while resonant Raman measurements revealed the emergence of new two-phonon modes highly sensitive to composition \cite{yetkin2024a}. Ramanandan et al. investigated secondary oxide-based phases to better understand the formation of BaZrS$_3$ from sulfurized oxide precursors \cite{ramanandan2023a}, while Pandey et al. studied the temperature dependence of phonon frequencies and compared their results with density functional theory (DFT) calculations \cite{pandey2020a}. Several other studies have also explored the vibrational properties of BaZrS$_\text{3}$ through Raman spectroscopy \cite{romagnoli2024a, yu2021a, gross2017a, kayastha2023a, vincent2023a}, yet fundamental questions remain about its lattice dynamics and electron-phonon interactions. Despite these efforts, a comprehensive peak identification and symmetry assignment from high-quality single crystals remains lacking. Such analysis is essential for a deeper understanding of lattice dynamics and electron-phonon coupling in BaZrS$_\text{3}$, which directly impacts its optoelectronic properties. A thorough analysis of a high-quality crystal can also establish Raman spectroscopy as a promising technique for defect characterization in this compound \cite{dimitrievska2019a}. 

In this work, we grew a high-quality BaZrS$_\text{3}$ single crystal and performed a comprehensive investigation of its optoelectronic and vibrational properties. Using power-dependent PL spectroscopy, we demonstrate a strong emission dominated by band-to-band transitions, featuring a single, narrow peak -- significantly sharper than previously reported spectra -- and provide insights into the carrier dynamics using time-resolved PL (TRPL) measurements. To determine the properties of the phonons, we use multiwavelength excitation Raman spectroscopy to exploit resonance effects and polarization-dependent Raman spectroscopy to assign symmetries to all 23 theoretically predicted Raman modes. These experimental results are further supported by first-principles lattice dynamics calculations, offering a more detailed understanding of phonon modes and electron-phonon interactions. Our findings establish a more definitive picture of the intrinsic optoelectronic and vibrational characteristics of BaZrS$_3$, helping to resolve discrepancies in prior reports and highlighting its potential as well as limitations for optoelectronic applications.

\section{Methods}

\small

We grew single crystals of BaZrS$_\text{3}$ using the chemical vapor transport (CVT) technique. The starting materials -- BaS (Sigma Aldrich, 99.9\%), Zr (Cerac, 99.8\%), and S (Sigma Aldrich, 4N8) -- were weighed in stoichiometric proportions for BaZrS$_\text{3}$ and loaded into a glassy carbon boat. The boat was placed inside a quartz ampoule with a diameter of 28 mm and a length of 220 mm. Iodine was added as a transport agent at a concentration of 5 mg/cm$^\text{3}$. The ampoule was evacuated, sealed, and positioned in a two-zone furnace. The temperature was gradually increased from 950$^\circ$C to 1050$^\circ$C to initiate the reaction between the starting materials. The ampoule was maintained under a temperature gradient from 1150$^\circ$C (hot zone) to 1050$^\circ$C (cold zone) for 240 hours, facilitating material transport from the hot to the cold end. During this growth process, an elongated, needle-like crystal, measuring 4 mm in length, grew at the center. After the furnace cooled to ambient temperature, the ampoule was opened, allowing volatile gaseous by-products to evaporate. Finally, the crystal was rinsed with ethanol to remove soluble, moisture-sensitive residues. In contrast to previous studies that employed broader temperature ranges or shorter reaction times -- often resulting in polycrystalline material or secondary phases such as BaS$_\text{3}$, ZrS$_\text{3}$, or ZrO$_\text{2}$ \cite{pradhan2023a, agarwal2025a} -- we used a more precisely controlled CVT process with a narrower temperature gradient and an extended reaction duration. This optimized protocol enabled the growth of high-quality, phase-pure BaZrS$_\text{3}$ single crystals, as confirmed by XRD and Raman spectroscopy.

The crystal structure of nominally BaZrS$_\text{3}$ was determined using single crystal X-ray diffraction (XRD). Single crystal XRD measurements were performed using a Bruker APEX-II CCD diffractometer with Mo-K$\alpha$ radiation ($\lambda$ = 0.71073 Å). Data acquisition was carried out using Bruker APEX3, while cell refinement and data reduction were performed using Bruker SAINT. A multiscan absorption correction was applied using Bruker TWINABS. Structure solution and refinement was performed with SHELXT \cite{sheldrick2015b} and SHELXL2018/3.3 \cite{sheldrick2015a}. The refinement resulted in a distorted perovskite-type structure with the orthorhombic space group \textit{Pnma}. The key refinement parameters are $a=7.076(3)$ Å, $b=9.991(4)$ Å, $c=7.027(3)$ Å, $V = 496.8(3)$ Å$^\text{3}$; $Z = 4$; $\mu = 11.043$ mm$^\text{-1}$; $T = 296(2)$ K; measurement range: $0\leq h \leq 10$, $0\leq k \leq 14$, $0\leq l \leq 10$, $4.079^\circ\leq \theta \leq 32.027^\circ$; 16493 observed reflections (for both individua), 2507 reflections after merging in point group \textit{mmm} (2335 independent reflections with $I > 2\sigma\left(I\right)$), completeness: 99.8 \%; $R_\text{int} = 0.038$, $R_\sigma = 0.026$; refined parameters: 30; $R_1$ (all data) = 0.0358, $wR_\text{2}(F^\text{2}) = 0.0799$, $\text{goodness-of-fit (GoF)} = 1.061$; $-1.46 \leq \Delta p \leq 2.24$. The crystal was refined as a non-merohedral twin using a HKLF5 file produced by TWINABS with 2 individua of $\approx$ 30 \% and 70 \% relative intensities.

X-ray fluorescence (XRF) measurements were performed using a Bruker M4 Tornado Micro-XRF spectrometer equipped with a Rh excitation beam and two detectors. To ensure reliable results, the system was calibrated with elemental standards. The spot size was 20 mm. High compositional accuracy was achieved by averaging data from 20 local measurement points within a $3\times3$ mm$^\text{2}$ square area.

Raman and photoluminescence (PL) spectra were measured using a WITec alpha300 R and Horiba LabRam confocal Raman microscope in a backscattering configuration. Multiwavelength excitation Raman measurements were conducted using 488, 532, 638, and 785 nm lasers. The excitation laser was focused onto the sample surface using a long-range 50x microscope objective with a numerical aperture of 0.55, yielding beam diameters approximately 1.1 $\mu$m for the 488 nm laser, 1.2 $\mu$m for the 532 nm laser, and 1.7 $\mu$m for the 785 nm laser. To ensure that the selected laser power did not induce heating effects or structural modifications in the sample, a power-dependent study was performed. The Raman spectrum was recorded at a fixed location on the sample while systematically increasing the laser power density, starting from the lowest available setting. Each spectrum was evaluated for shifts in peak positions, broadening of peaks, or the emergence of additional features. The maximum power level at which no such changes were detected was chosen as the optimal power setting for all subsequent measurements. The backscattered light was analyzed using two spectrometers. For 488, 532 and 638 nm excitation, a spectrometer equipped with a holographic grating (150 g/mm for PL, 1800 g/mm for Raman) and a thermoelectrically cooled CCD was used. For measurements with 785 nm excitation, a spectrometer with a 1200 g/mm grating and a deep-depletion cooled CCD was employed. All Raman measurements were performed using a power density of 20 mW/cm$^\text{2}$, with each spectrum acquired over 60 seconds and averaged over three acquisitions to improve the signal-to-noise ratio. Raman peak positions were calibrated against the Si reference peak at 520 cm$^\text{-1}$. All measurements were conducted at room temperature.

Time-resolved photoluminescence measurements were performed using a MicroTime 100 time-resolved confocal microscope coupled with a PicoQuant detection unit. Photocarriers were excited using a pulsed laser at $\lambda = 639$ nm, with a pulse duration of $<100$ ps, a beam diameter of 130 $\mu$m, and a repetition rate of 64 MHz. The excitation laser was focused onto the sample surface through a long-range 20x objective with a numerical aperture of 0.45. Photons emitted from the sample were collected through the objective and guided via a 50 $\mu$m diameter optical fiber to the detector unit, which comprised a FluoTime 300 photospectroscopmeter, a monochromator and an InGaAs-based photomultiplier detector (950–1450 nm) from Hamamatsu. A 667 nm cutoff filter was employed to prevent the collection of reflected photons from the excitation laser. The uncertainty of the extrapolated decay time is given as the instrumental resolution limit of 200 ps.

Raman calculations were carried out using AiiDAlab-QE, a Quantum ESPRESSO-based web application \cite{aiidalab_qe}, in combination with the vibroscopy external plugin (AiiDAlab-qe-vibroscopy) \cite{aiidalab_qe_vibroscopy_github, yakutovich2021a}. This plugin integrates automated AiiDA-based workflows for phonon and Raman calculations by employing a finite-displacement and finite-field approach (aiida-vibroscopy) \cite{bastonero_automated_2024}. The DFT calculations were performed using Quantum ESPRESSO 7.2 \cite{Giannozzi_2009, Giannozzi_2017}. Before computing the vibrational spectra, the atomic positions and cell parameters were optimized using the PBE functional \cite{perdew_Generalized_1996} until the total energy and atomic forces were converged below $5 \times 10^{-6}$ Ry/atom and $5 \times 10^{-5}$ Ry/Bohr, respectively. Norm-conserving pseudopotentials from the Pseudo-Dojo library were employed \cite{pseudodojo}. A plane-wave cutoff of 70 Ry was used for the wavefunctions, while a cutoff of 280 Ry was applied for the charge density. The Brillouin zone was sampled using a uniform $k$-point grid with a spacing of 0.1 Å$^\text{-1}$, resulting in a $9\times9\times7$ mesh for the primitive unit cell containing 20 atoms.

\normalsize

\section{Results}

\begin{figure}[t!]
    \centering
    \includegraphics[width=0.8\columnwidth,trim={0 0 0 0},clip]{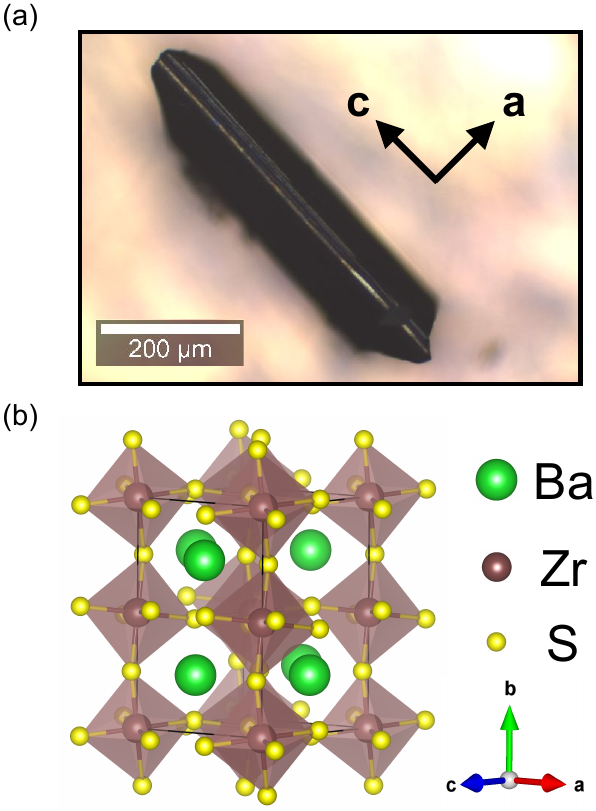}
    \caption{(a) Optical microscope image of the needle-like single crystal studied in this work. (b) Crystal structure of BaZrS$_\text{3}$, generated in VESTA from the CIF file obtained through single-crystal XRD refinement.}
    \label{fig:Figure1}
\end{figure}

First, we perform a chemical and structural characterization of the synthesized single crystal. Fig. \ref{fig:Figure1}(a) presents an optical microscopy image of the needle-like single crystal of BaZrS$_\text{3}$. The chemical composition of the crystal was investigated by X-ray fluorescence spectroscopy, revealing a Ba/Zr ratio of $0.93 \pm 0.04$. This observation suggests the possibility of slight off-stoichiometry, which may be accommodated by defect formation. However, as detailed later, XRD and Raman spectroscopy confirms the absence of secondary phases, indicating that any deviations from stoichiometry are not associated with macroscopic phase separation. Given the relatively high formation energies of isolated barium vacancies ($V_\text{Ba}$) and cation antisites ($\text{Ba}_\text{Zr}$, $\text{Zr}_\text{Ba}$) predicted by DFT studies \cite{yuan2024a, desai2025a}, these defects are unlikely to exist in significant concentrations, even under barium-poor conditions. Instead, we expect charge-neutral defect complexes -- such as $\text{Zr}_\text{Ba}^{2+} + \text{Ba}_\text{Zr}^{2-}$ or $\text{Zr}_\text{Ba}^{2+} + V_\text{Ba}^{2-}$ -- to form preferentially, as these would maintain charge neutrality and minimize the electrostatic energy of the lattice. Although such complexes have not yet been explicitly investigated in BaZrS$_\text{3}$ through DFT calculations, their formation is strongly supported by the defect chemistry in related chalcogenide materials, such as Cu(In,Ga)S$_\text{2}$ and kesterites \cite{park2018a, schorr2020a}. Considering the estimated 4\% experimental uncertainty in the XRF measurements, along with our comprehensive structural and optoelectronic characterization presented below, we conclude that the crystal is close to stoichiometric, with any native defects being low in concentration and likely existing as electrically benign complexes.

The structural analysis of the crystal confirms the distorted-perovskite type, orthorhombic crystal structure with corner-sharing ZrS$_\text{6}$ octahedra (space group \textit{Pnma}). Fig. \ref{fig:Figure1}(b) depicts the determined crystal structure, generated in VESTA from the CIF file obtained through single-crystal XRD structure determination. The obtained lattice parameters are $a=7.076(3)$ Å, $b=9.991(4)$ Å, and $c=7.027(3)$ Å. The atomic position parameters, determined from the structural analysis, are presented in Table \ref{tbl:AtomicPositions}. These parameters further confirm the precise positions of the constituent elements within the orthorhombic unit cell. The CIF has been deposited with the CCDC and assigned CSD Number 2465609.

\begin{table}[t!]
\small
\caption{Atomic position parameter of BaZrS$_\text{3}$ (space group \textit{Pnma}). The CIF has been deposited with the CCDC and assigned CSD Number 2465609}
\vspace{0.25cm}
\label{tbl:AtomicPositions}
\begin{tabular*}{\columnwidth}{@{\extracolsep{\fill}}lcccc}
     & Wyckoff position & x & y & z \\ \hline
    Ba & 4c & 0.9615 & 0.2500 & 0.4926 \\
    Zr & 4a & 0.0000 & 0.0000 & 0.0000 \\
    S1 & 4c & 0.0042 & 0.2500 & 0.9403 \\
    S2 & 8d & 0.7130 & 0.9694 & 0.2130 \\
    \hline 
\end{tabular*}
\end{table}

\begin{figure*}[t!]
    \centering
    \includegraphics[width=\textwidth,trim={0 0 0 0},clip]{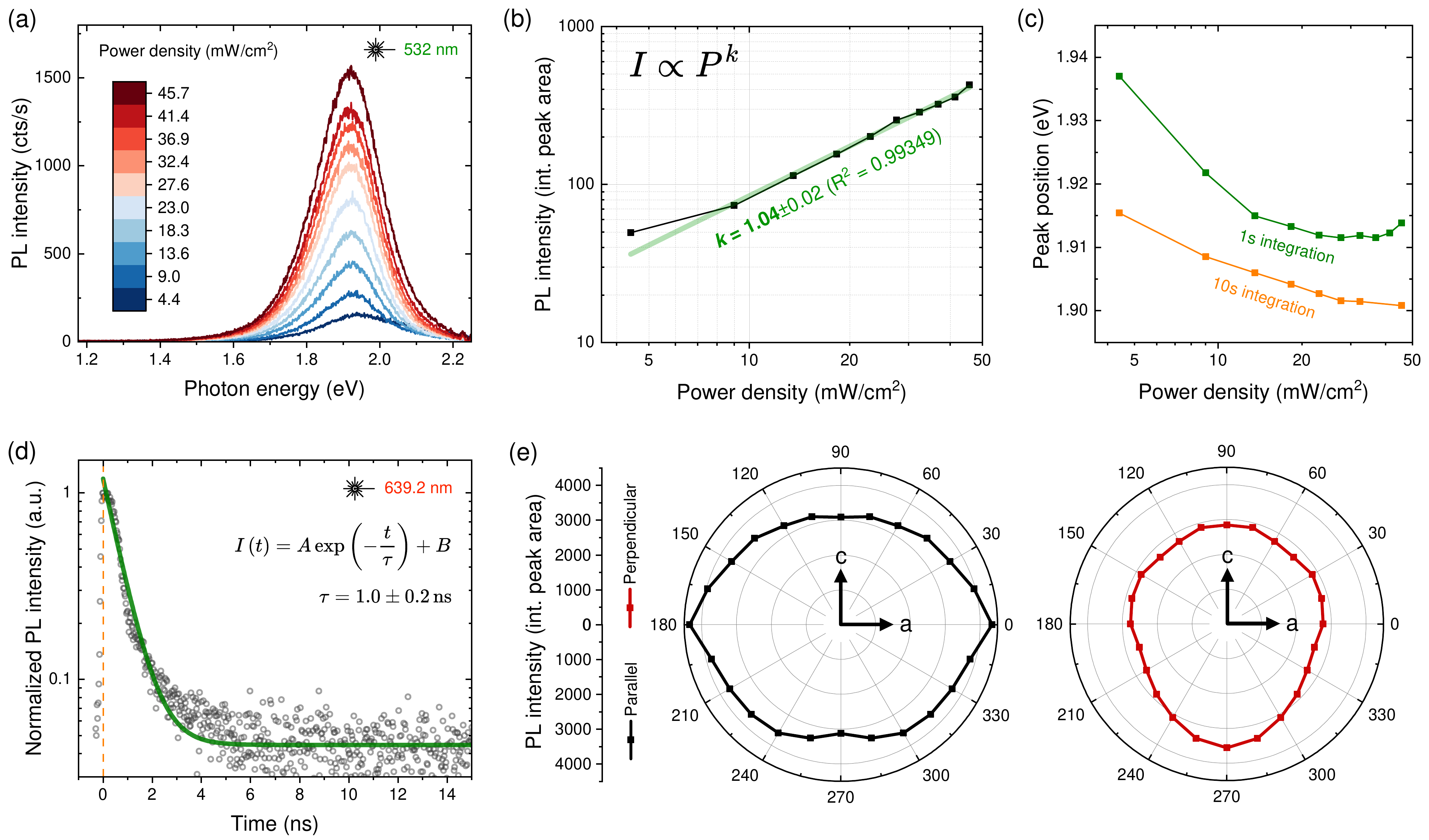}
    \caption{(a) Excitation power-dependent PL spectra of the BaZrS$_\text{3}$ single crystal measured at room temperature. (b) Integrated PL intensity, $I$, as a function of excitation power density, $P$. The fit follows a power-law relation with an exponent $k=1.04\pm0.02$, indicating a band-to-band-like transition. (c) PL peak position as a function of excitation power density. Two integration times were used to illustrate the effect of total energy dose on the peak position. (d) Time-resolved PL (TRPL) transient fitted with a monoexponential decay, yielding a characteristic decay time of $\tau=1.0 \pm 0.2$ ns. (e) Angle-resolved polarization-dependent PL intensities measured on the (010) plane, revealing weak optical anisotropy in the \textit{xy} plane.}
    \label{fig:Figure2}
\end{figure*}

\subsection*{Photoluminescence spectroscopy of BaZrS$_\text{3}$}

To investigate the optoelectronic properties of the BaZrS$_\text{3}$ single crystal, we performed photoluminescence (PL) spectroscopy at room temperature, studying the dependence on excitation power density and polarization, as well as the time-resolved PL decay.

Figure \ref{fig:Figure2}(a) shows the power-dependent PL spectra of the BaZrS$_\text{3}$ single crystal. Previous reports on the PL spectra of BaZrS$_\text{3}$ have been highly inconsistent, often featuring broad, asymmetric, and sometimes even multi-component emission bands \cite{wei2019a, comparotto2020a, niu2017a, pradhan2023a, ye2024a, yang2022a, surendran2021a, gupta2020a, m2021a, ye2022a}. In contrast, we observe a single symmetric emission peak in all our spectra, which can be fitted with a single Gaussian function. The peak width ranges from 0.19 to 0.24 eV at maximum power, which, to the best of our knowledge, is the narrowest emission peak reported for BaZrS$_\text{3}$. As the width of the PL peak is related to the Urbach energy and the density of shallow defects, this suggests that our BaZrS$_\text{3}$ crystal exhibits higher optoelectronic quality compared to previous reports. It is also worth noting that our measurements were conducted at room temperature, where Ye et al. recently reported no detectable PL from BaZrS$_\text{3}$ single crystals \cite{ye2024a}. We observe no emission features at lower photon energies, where Márquez et al. and Pradhan et al. have reported midgap defect peaks ($\sim$1.2 eV) attributed to undetected secondary phases and sulfur interstitials, which are hypothesized to quench PL in most BaZrS$_\text{3}$ samples \cite{m2021a, pradhan2024a}. Finally, we estimated the photoluminescence quantum yield (PLQY) to be approximately 0.005\% under 532 nm excitation, calibrated against an internal perovskite reference standard with a known PLQY of 0.03\% as reported by Lai et al. \cite{lai2022a}.

These discrepancies between our results and previous reports suggest that the multi-component features and weak PL intensities reported elsewhere are not intrinsic to BaZrS$_\text{3}$, but arise from synthesis-related imperfections, particularly in thin films. Given the high synthesis temperature of BaZrS$_\text{3}$, structural defects and impurities are more likely to become quenched-in, making defect control more critical for optimizing optoelectronic quality than in lead-halide perovskites, where charged point defects can be eliminated (or "self-healed") through ionic diffusion \cite{cahen2021a, finkenauer2022a}. However, our results demonstrate that despite the relatively rigid crystal structure of BaZrS$_\text{3}$, a single-component PL emission peak is achievable.

To investigate the nature of the radiative emission peak, the integrated PL intensities have been fitted to the power law $I \propto P^k$ in Figure \ref{fig:Figure2}(b), where $I$ is the integrated peak intensity, $P$ is the excitation power density, and $k$ is an exponent. When $k\leq1$, the transition is associated with defect recombination, whereas  $k\geq1$ typically indicates excitonic transitions \cite{schmidt1992a, levanyuk1981a}. For the BaZrS$_\text{3}$ single crystal studied here, we obtain $k=1.04\pm0.02$, suggesting that the radiative emission primarily arises from band-to-band transitions. This is a promising result for the material’s potential in optoelectronic applications.

We observe a shift in the PL emission peak with increasing integration time, as shown in Figure \ref{fig:Figure2}(c). While excitonic effects could, in principle, lead to transitions slightly below the bandgap energy, we believe this shift is more likely due to sample heating. A similar effect is observed when measuring the PL under 488 nm excitation, as shown in Figure S2 in the Supporting Information. Comparison of the PL spectra excited with 532 nm and 488 nm reveals a redshift of approximately 30 meV in the emission peak. While the spectral broadening is negligible, we attribute the redshift to local heating resulting from the smaller penetration depth of the 488 nm laser, as well as to the enhanced contribution of surface states, consistent with observations previously reported in halide perovskite single crystals \cite{chen2021a}.

Reported PL peak positions in the literature vary widely, ranging from 1.65 eV to 1.95 eV \cite{agarwal2025a}. However, given the strong PL intensity, narrow peak width, and band-to-band-like transition observed in our single crystal, we propose that the peak position of 1.92$\pm$0.01 eV represents the intrinsic bandgap of BaZrS$_\text{3}$ at room temperature. Yuan et al. \cite{yuan2024a} recently demonstrated that several intrinsic defects form shallow donor levels with relatively low formation enthalpies. In addition to chemical impurities and structural defects, these may contribute to the broad range of reported emission peak positions and peak asymmetries in the literature. The observed peak shift with increasing power density, as well as the influence of total injected energy — illustrated here by the shift with increasing integration time during the data acquisition — underscores the importance of reporting such measurement conditions when comparing PL data across different studies.

The TRPL transient of the the BaZrS$_\text{3}$ single crystal is shown in Figure \ref{fig:Figure2}(d). The transient is well described by a monoexponential decay, yielding a characteristic decay time of $\tau= 1.0\pm0.2$ ns. This result is promising, given that the crystal surfaces have not been passivated. The TRPL decay time is comparable to that reported for single crystals by Zhao et al. \cite{zhao2024a}, and for epitaxially grown BaZrS$_\text{3}$ thin films by Surendran et al. \cite{surendran2021a}, where the steady-state PL spectrum also exhibits a single, symmetric emission peak. Furthermore, similar decay times were obtained by transient absorption spectroscopy on epitaxially grown BaZrS$_\text{3}$ thin films \cite{you2024a}.

Longer TRPL decay times have been observed by Ye et al. using single crystals \cite{ye2022a}, and Wei et al. with thin films synthesized by sulfurization of PLD-deposited oxide precursors \cite{wei2019a}. However, these longer decay times are believed to originate from carrier localization, likely due to defect trapping, as indicated by biexponential decay behavior. The presence of multiple recombination channels in these samples aligns with the more complex steady-state PL spectra, which often feature several emission peaks \cite{ye2024a}. While Surendran et al., Wei et al., and Ye et al. all report biexponential decays, the shorter decay components in their studies are comparable to our characteristic decay of $\tau\sim1$ ns. Yuan et al. \cite{yuan2024a} calculated an upper limit of 33 ns for the non-radiative lifetime in BaZrS$_\text{3}$ synthesized under sulfur-rich conditions. Ye et al. initially reported decay times of up to 50 ns \cite{ye2022a}, but later clarified that these long lifetimes are likely associated with carrier localization through defect trapping, which explains the coexistence of very long TRPL decay times and low PL quantum yields \cite{ye2024a}.

Finally, we performed angle-resolved polarization-dependent PL spectroscopy on the (010) plane, as shown in Figure \ref{fig:Figure2}(e). While some degree of optical anisotropy is observed in the \textit{xy} plane, it is far less pronounced than in BaTiS$_3$, which adopts a quasi-1D structure believed to drive the giant optical anisotropy in that compound \cite{yang2023a}. Although the emission characteristics appear largely isotropic, the slight anisotropy observed still warrants further exploration of optical anisotropy in other crystal planes of BaZrS$_\text{3}$.

\subsection*{Vibrational properties and phonons in BaZrS$_\text{3}$}

Group theory analysis predicts 60 phonon modes for BaZrS$_\text{3}$ (space group \textit{Pnma} (No. 62), point group mmm) with the following irreducible representations \cite{gallego2019a}:

\begin{equation}
    \begin{split}
        \Gamma_\mathrm{vib} = & \, 7\mathrm{A}_{\mathrm{g}}+8\mathrm{A}_\mathrm{u}+5\mathrm{B}_{1\mathrm{g}}+10\mathrm{B}_{1\mathrm{u}} \\
        & + 7\mathrm{B}_{2\mathrm{g}}+8\mathrm{B}_{2\mathrm{u}}+5\mathrm{B}_{3\mathrm{g}}+10\mathrm{B}_{3\mathrm{u}}
    \end{split}
\end{equation}

\noindent where (i)  $\text{B}_{1\text{u}} + \text{B}_{2\text{u}}+\text{B}_{3\text{u}}$  are acoustic modes, (ii) $9\text{B}_{1\text{u}}+7\text{B}_{2\text{u}}+9\text{B}_{3\text{u}}$ are infrared active modes, and (iii) $7\text{A}_{\text{g}}+5\text{B}_{1\text{g}}+7\text{B}_{2\text{g}}+5\text{B}_{3\text{g}}$ are Raman active modes. Here, A and B denote non-degenerate symmetric and asymmetric modes with respect to the principal rotational axis (under all 180$^\circ$ rotations in the point group), while the subscripts $\text{g}$ and $\text{u}$ indicate mode symmetry and asymmetry under inversion, respectively. The numerical subscripts in the phonon modes denote asymmetry with respect to specific mirror planes: "1" corresponds to asymmetry with respect to the \textit{yz} plane, "2" to the \textit{xz} plane, and "3" to the \textit{xy} plane.

To determine the properties of these phonon modes, we performed multiwavelength excitation Raman measurements in both non-polarized and polarized configurations, supported by phonon and Raman calculations in DFT.

\begin{figure}[ht!]
    \centering
    \vspace{0.45cm}
    \includegraphics[width=0.97\columnwidth,trim={0 0 0 0},clip]{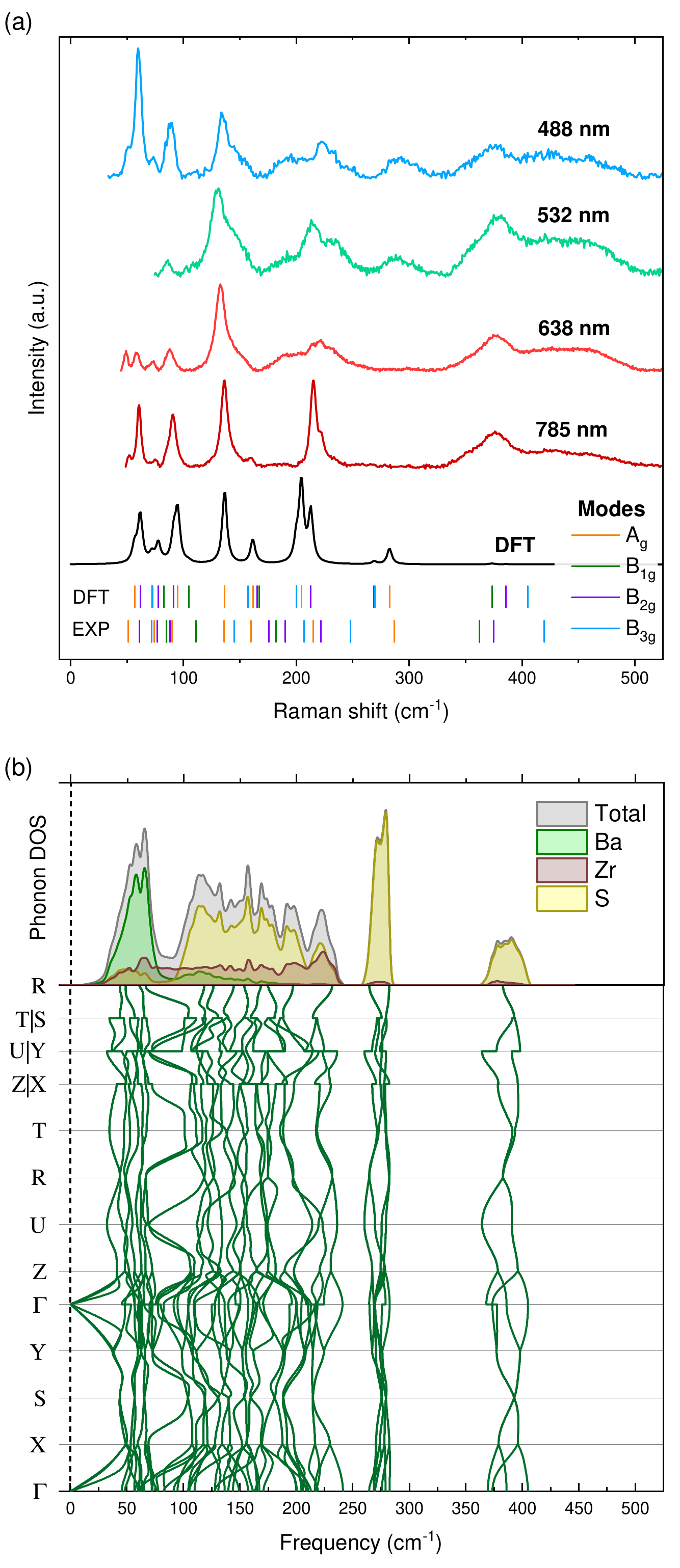}
    \vspace{0.25cm}
    \caption{(a) Non-polarized Raman spectra of BaZrS$_\text{3}$ measured with 488, 532, 638 and 785 nm laser excitation, along with the DFT calculated Raman spectrum. Vertical lines beneath the spectra show a comparison between experimentally determined Raman peak positions (labeled "Exp"), obtained through deconvolution, and those derived from DFT (labeled "DFT"). (b) Phonon density of states and DFT-calculated phonon dispersion along high-symmetry directions of BaZrS$_\text{3}$.}
    \label{fig:Figure3}
    \vspace{-0.75cm}
\end{figure}

\subsection*{Multiwavelength excitation Raman measurements of BaZrS$_\text{3}$}

Non-polarized Raman spectra of BaZrS$_\text{3}$, measured using 488, 532, 638, and 785 nm laser excitations, along with the non-resonant DFT-calculated Raman spectrum, are shown in Figure \ref{fig:Figure3}(a).

The Raman spectra obtained with 488 and 532 nm excitations exhibit very similar features, whereas those measured with 638 and 785 nm excitations show differences in the intensity distribution of the Raman peaks. These variations arise from differences in laser penetration depth and resonance conditions. Based on the absorption spectrum of BaZrS$_\text{3}$ \cite{nishigaki2020a}, the penetration depth ranges from approximately 10 µm for a 785 nm wavelength to 1 µm for 638 nm and around 100 nm for 532 and 488 nm. This suggests that measurements with shorter wavelength excitations (488 and 532 nm) are more sensitive to surface effects, such as surface roughness or the formation of a native oxide layer, as reported by Agarwal et al. \cite{agarwal2024a}. These surface effects reduce the intensity and broaden the Raman peaks. In contrast, the 785 nm excitation probes a larger volume of material, resulting in narrower and more intense Raman peaks, as seen in Figure \ref{fig:Figure3}(a). Additionally, since our PL measurements indicate a direct bandgap of $1.92\pm0.01$ eV for BaZrS$_\text{3}$, the 638 nm excitation (1.94 eV) is resonant. This resonance enhances specific Raman modes and modifies the spectral shape, particularly in the lower-frequency region of Figure \ref{fig:Figure3}(a).

We note that all observed peaks in the Raman spectra of the single crystal can be attributed to the BaZrS$_\text{3}$ phase, with no evidence for the presence of secondary phases such as ZrS$_\text{2}$, BaS, BaS$_\text{3}$, BaZrO$_\text{3}$, ZrO$_\text{2}$, or oxysulfides, based on comparison with reference Raman spectra reported in the literature \cite{ma2016a, jaroudi1999a, ramanandan2023a}.

The non-resonant DFT calculated Raman spectrum shows good agreement with the experimental data, particularly when comparing the Raman intensity profile and peak positions to the spectrum measured with 785 nm excitation. This is expected, as the 785 nm spectrum best reflects the bulk characteristics of the BaZrS$_\text{3}$ crystal. Slight differences in the Raman peak positions between the experimental and DFT spectra can be observed in the high-frequency region, around 200 cm$^\text{-1}$. These differences are attributed to anharmonic effects, which are more pronounced for lighter sulfur atoms due to their larger vibrational amplitudes. This leads to an underestimation of frequencies in the harmonic approximation. Additionally, electron-phonon interactions, which are not fully captured in standard DFT calculations, can further renormalize the vibrational modes, contributing to the observed deviation between theoretical and experimental spectra.

A more detailed analysis of the phonon nature in BaZrS$_\text{3}$ can be obtained from the calculated phonon dispersion along high-symmetry directions of the Brillouin zone, as shown in Figure \ref{fig:Figure3}(b), along with the elemental phonon density of states (PDOS). Several distinct regions can be identified in the phonon dispersion diagram: (i) the low-frequency region ($<\text{100}$ cm$^\text{-1}$), predominantly dominated by Ba-related vibrations with smaller contributions from Zr atoms, (ii) the intermediate region (100 – 240 cm$^\text{-1}$), which corresponds to mixed contributions from Zr- and S-related vibrations, and (iii) two high-frequency regions (260 – 280 cm$^\text{-1}$ and 370 – 420 cm$^\text{-1}$), primarily attributed to S-related vibrations. Additionally, two phonon gaps are observed: one in the 240 – 260 cm$^\text{-1}$ frequency region, and another in the 280 – 370 cm$^\text{-1}$ region. These gaps arise from the mass differences between Ba, Zr, and S atoms, as well as the strong bonding within the ZrS$_\text{6}$ octahedra compared to the weaker interactions involving Ba. The phonon gaps restrict available phonon scattering channels, which in turn enhances the energy transfer from excited carriers to the lattice, promoting non-radiative recombination. While the existence of phonon gaps in BaZrS$_\text{3}$ suggests enhanced electron-phonon coupling and an increased probability of defect trapping, in agreement with the observations by Ye et al. \cite{ye2024a}, recent theoretical studies \cite{li2022a} have shown that electron-phonon coupling in BaZrS$_\text{3}$ is notably strong, significantly influencing polaron formation and carrier dynamics. The observed phonon gaps are consistent with a lattice that supports moderate to strong electron-phonon coupling, where specific vibrational modes preferentially interact with charge carriers, leading to polaronic effects. This behavior contrasts with more rigid semiconductors such as GaAs, Si, or GaN, where stronger bonding, lower dielectric polarizability, and higher phonon frequencies suppress such interactions. Therefore, the combination of a soft lattice, a polarizable bonding environment, and the presence of phonon gaps provides evidence for nontrivial electron-phonon interactions in BaZrS$_\text{3}$, which are crucial for understanding its carrier recombination and transport properties.

\begin{figure}[t!]
    \centering
    \includegraphics[width=0.9\columnwidth,trim={0 0 0 0},clip]{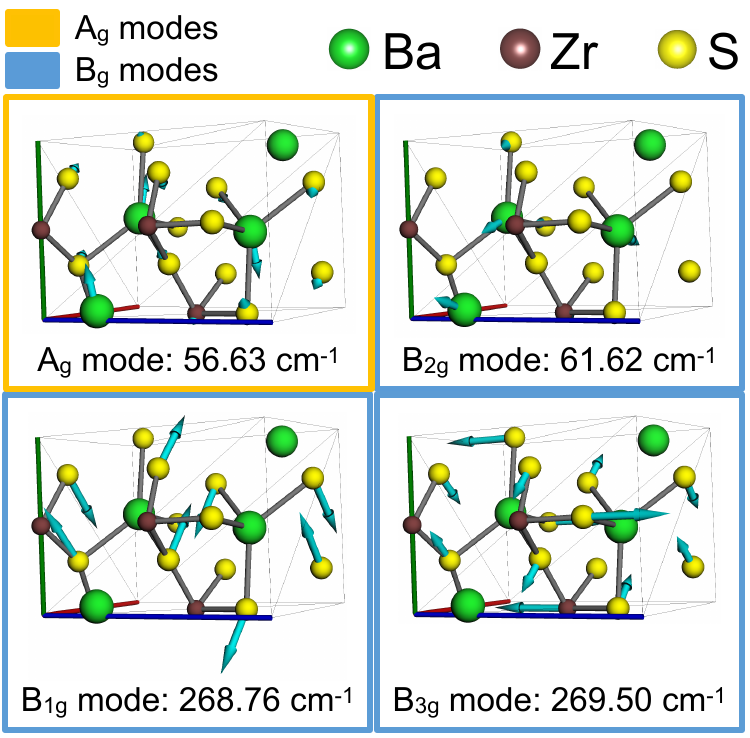}
    \caption{Calculated phonon displacements for Raman modes of BaZrS$_\text{3}$. Mode symmetries and frequencies (in cm$^\text{-1}$) are listed under each picture.}
    \label{fig:Figure6}
\end{figure}

Atomic displacements of the Raman modes were calculated to visualize the corresponding atomic motions. Figure \ref{fig:Figure6} shows the vibrational patterns of representative Raman-active modes, while the full set of mode representations is provided in Figure S5 in the Supporting Information. 

\begin{figure*}[t!]
    \centering
    \includegraphics[width=0.72\textwidth,trim={0 0 0 0},clip]{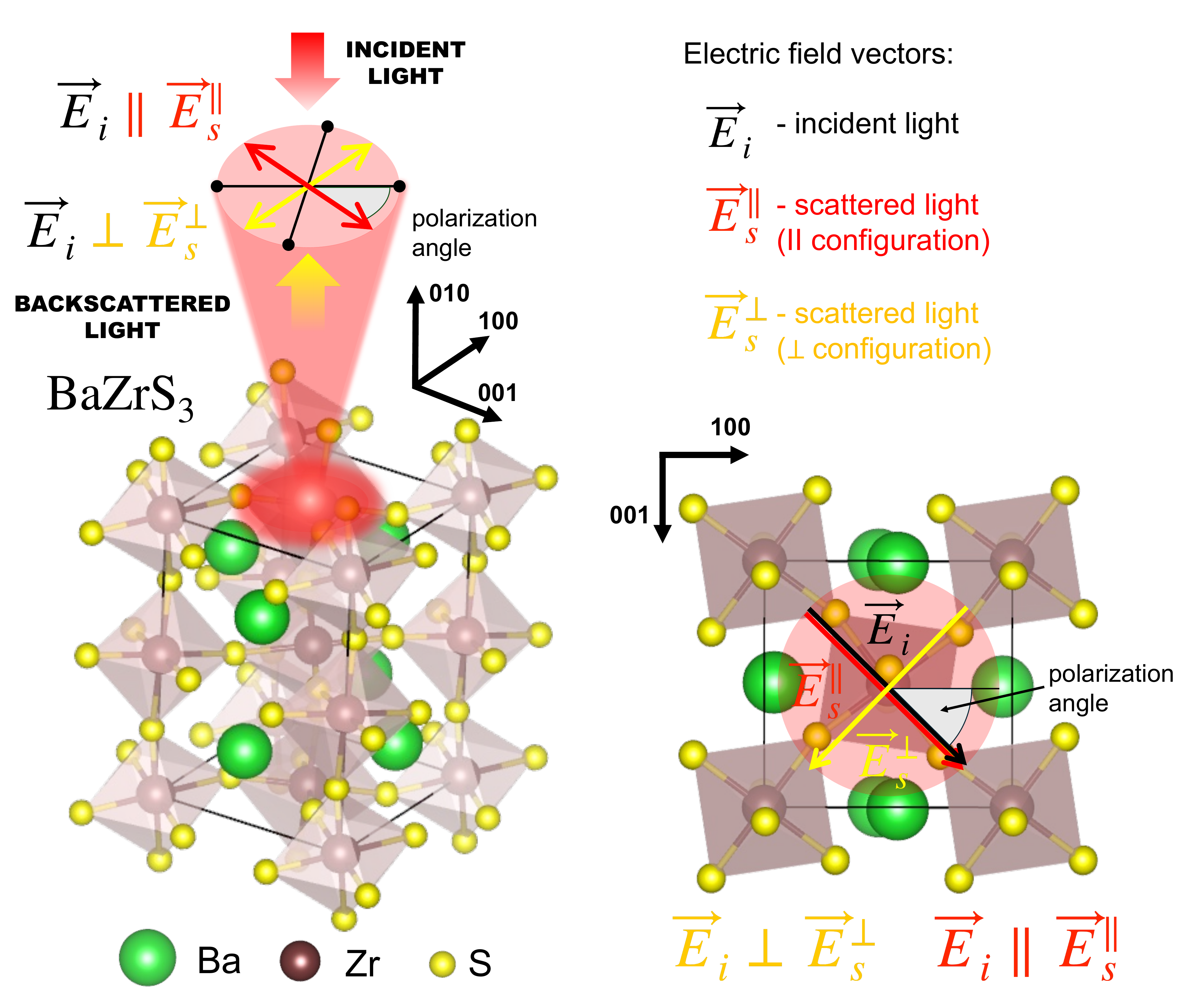}
    \caption{Schematic illustration of the Raman scattering polarization measurements. Raman measurements are performed on (010) basal plane with incident laser light (wide red arrow) projected along the crystallographic y-axis ([010] direction). The incident light polarization and the output polarization in perpendicular configuration are represented by the thin red and yellow arrows, respectively.}
    \label{fig:Figure5}
\end{figure*}

To determine the frequency and phonon order (one- or multi-phonon) of the Raman modes, we performed a deconvolution of the Raman spectra shown in Figure \ref{fig:Figure3}(a). The deconvolution was carried out simultaneously for all four spectra, allowing the identification of 23 one-phonon Raman modes, as predicted by theory. Each peak was modeled using a Lorentzian curve, characterized by its peak position, width, and intensity. The initial parameters for the one-phonon modes were derived from the simulated Raman spectra obtained from DFT calculations, while multi-phonon modes were introduced as needed during the fitting process. Due to the large number of variables involved, strict constraints were applied to prevent correlations among the parameters and ensure meaningful results. The peak positions and widths of the same Raman modes were tied together across all four spectra, while the intensities were treated as free parameters. Since peak widths are primarily determined by phonon lifetimes, which depend on the crystal quality, modes of the same phonon order are expected to have similar widths. Additionally, multi-phonon modes are expected to be broader compared to one-phonon modes. This approach resulted in a narrow range of allowed variations in the peak widths throughout the deconvolution process, enabling an unambiguous determination of the Raman mode frequencies. The representative deconvolution results for the Raman spectra measured with 488 nm excitation are shown in Figure S4 in the Supporting Information.

\subsection*{Angle-resolved Raman polarization measurements of BaZrS$_\text{3}$}

\begin{figure*}[ht!]
    \centering
    \includegraphics[width=0.9\textwidth,trim={0 0 0 0},clip]{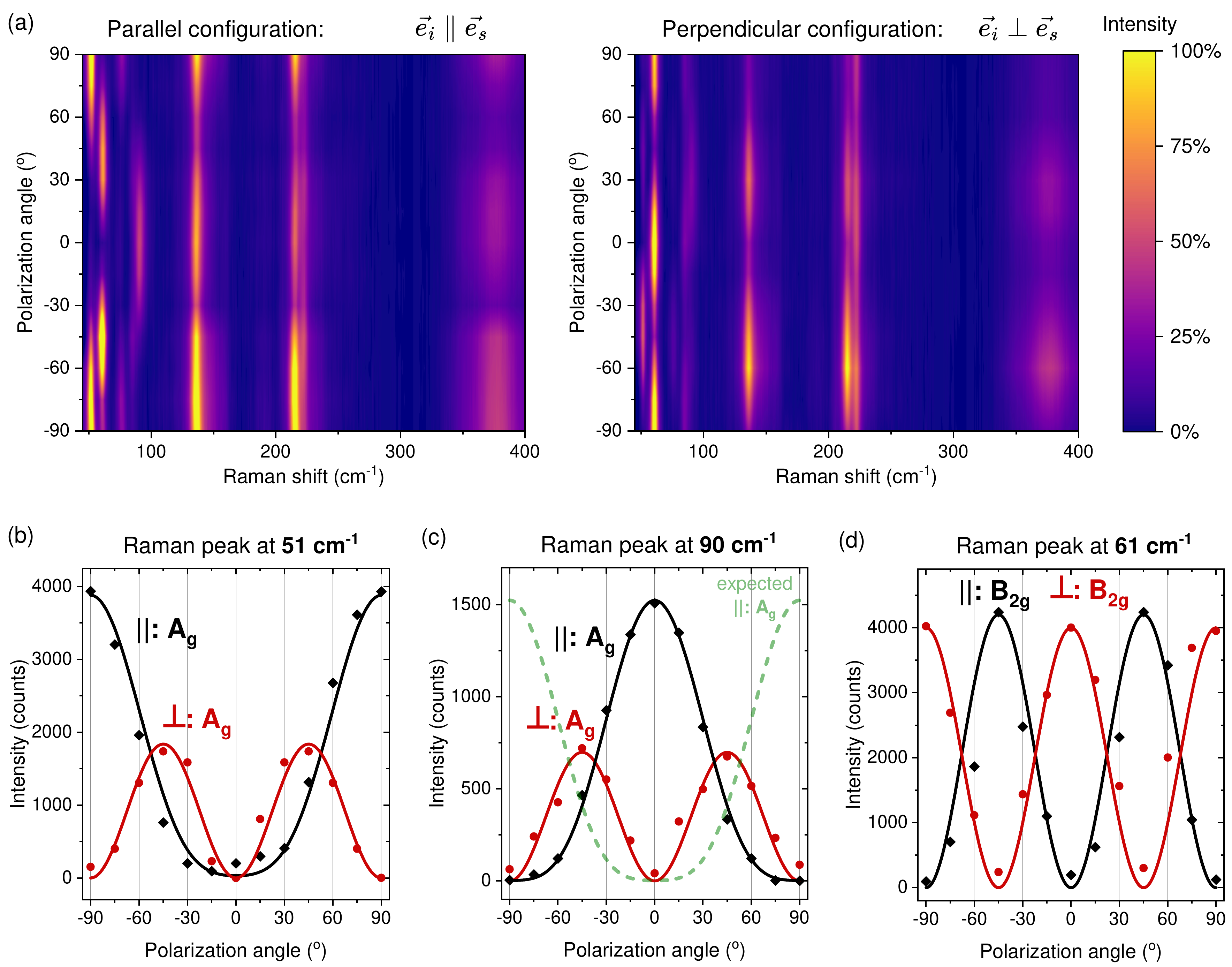}
    \caption{(a) Polarization Raman spectra of BaZrS$_\text{3}$ measured for various polarization angles ($\theta$) in perpendicular and parallel configuration. The spectra were acquired at room temperature and using 785 nm excitation wavelength. Both configurations were measured under the same experimental conditions. Polarization-dependent Raman intensities for two characteristic Raman peaks centered at (b) 51, (c) 90, and (d) 61 cm$^\text{-1}$, showing the modeling of the experimental data according to the equations presented in Table \ref{tbl:TableWithEquations}. Dot and diamond symbols represent the experimental data, while full lines present the fits. The standard uncertainty in the intensity is considered to be $\pm$2\%.}
    \label{fig:Figure4}
\end{figure*}

As the symmetry of the Raman modes cannot be directly determined from multiwavelength Raman measurements, we performed angle-resolved polarization Raman measurements.

Under non-resonant conditions, the Raman tensors for the Raman-active phonon modes are defined as:

\begin{equation}
    \begin{split}
        \mathfrak{R}\left(\mathrm{A_g}\right) &=
        \begin{pmatrix}
            a & 0 & 0 \\
            0 & b & 0 \\
            0 & 0 & c
        \end{pmatrix}, \quad
        \mathfrak{R}\left(\mathrm{B_{1g}}\right) =
        \begin{pmatrix}
            0 & d & 0 \\
            d & 0 & 0 \\
            0 & 0 & 0
        \end{pmatrix}, \\
        \mathfrak{R}\left(\mathrm{B_{2g}}\right) &=
        \begin{pmatrix}
            0 & 0 & e \\
            0 & 0 & 0 \\
            e & 0 & 0
        \end{pmatrix}, \quad
        \mathfrak{R}\left(\mathrm{B_{3g}}\right) =
        \begin{pmatrix}
            0 & 0 & 0 \\
            0 & 0 & f \\
            0 & f & 0
        \end{pmatrix}
    \end{split}
\end{equation}

\noindent where $a$, $b$, $c$, $d$, $e$ and $f$ are the Raman tensor coefficients (elements). 

The scattering intensity of each mode in the Raman spectra is given by:

\begin{equation}
    I_S \propto C\left(\omega_p\right) \left| \vec{e_i} \cdot \mathfrak{R} \cdot \vec{e_s} \right|^2
\end{equation}

\noindent where $\vec{e_i}$ and $\vec{e_s}$ are the unitary polarization vectors of the incident light and backscattered light, respectively. The coefficient $C\left(\omega_p\right)$ describes the dependence of the Raman mode intensity from the phonon frequency $\omega_p$ and the incident laser frequency $\omega_i$. It is defined as:

\begin{equation}
    C\left(\omega_p\right) = \frac{\omega_i\left(\omega_i-\omega_p\right)^3}{\omega_p\left[1-\exp{\left(-\hbar\omega_p/k_\mathrm{B}T\right)}\right]}
\end{equation}

\noindent where $\hbar=h/2\pi$, with $h$ being the Planck constant, $k_\mathrm{B}$ is the Boltzmann constant and $T$ is the temperature. 

\begin{table*}[ht!]
\small
\caption{Angular dependencies of Raman modes intensity for BaZrS$_\text{3}$, in parallel and perpendicular measurement configurations on (010) basal plane, with $\theta$ being the polarization angle between $\vec{e_i}$ and (100).}
\vspace{0.25cm}
\label{tbl:TableWithEquations}
\begin{tabular*}{0.8\textwidth}{@{\extracolsep{\fill}}ccc}
    Phonon & Parallel & Perpendicular \\
    mode & configuration & configuration \\
    \hline \vspace{-0.3cm} \\
    A$_\mathrm{g}$ & $C\left(\omega_p\right) \times  \left(a \times \cos^2{\theta} + c \times \sin^2{\theta} \right)^2$ & $C\left(\omega_p\right) \times  \left(a - c\right)^2 \times \cos^2{\theta} \times \sin^2{\theta}$ \\
    B$_\mathrm{1g}$ & 0 & 0\\
    B$_\mathrm{2g}$ & $C\left(\omega_p\right) \times  \left| e \times \sin{2\theta} \right|^2$ & $C\left(\omega_p\right) \times  \left| e \times \cos{2\theta} \right|^2$ \\
    B$_\mathrm{3g}$ & 0 & 0 \vspace{0.1cm}  \\
    \hline 
\end{tabular*}
\end{table*}

Raman measurements were performed on the single crystal with the (010) lattice plane as the basal plane, as shown in Figure \ref{fig:Figure5}. Considering this configuration, the incident $\vec{e_i}$ and backscattering $\vec{e_s}$ polarization unit vectors can be defined as: 

\begin{equation}
    \begin{split}
        \vec{e_i} = \begin{pmatrix}
            \sin{\theta} & 0 & \cos{\theta}
        \end{pmatrix} \hspace{1.6cm}\\
        \vec{e}_s^{\,\parallel} = 
        \begin{pmatrix}
            \sin{\theta} \\
            0 \\
            \cos{\theta}
        \end{pmatrix}, \quad \text{and} \quad 
        \vec{e}_s^{\,\perp} = 
        \begin{pmatrix}
            \cos{\theta} \\
            0 \\
            -\sin{\theta}
        \end{pmatrix}
    \end{split}
\end{equation}

\noindent where $\theta$ is the polarization angle with respect to the [100] axis, and the symbols $\parallel$ and $\perp$ denote parallel and perpendicular polarization, respectively.

Combining equations (2-5) yields the Raman intensity dependency on the polarization angle for different phonon modes, which are presented in Table \ref{tbl:TableWithEquations}. We find that changes in the polarization angle result in selective activation or cancellation of phonon modes. When polarization Raman measurements are performed with the incident light onto the (010) basal plane (Figure \ref{fig:Figure5}), the B$_\text{1g}$ and B$_\text{3g}$ modes are silent, while the A$_\text{g}$ and B$_\text{2g}$ modes vary according to the polarization configuration. 

Figure \ref{fig:Figure4} presents the experimental polarization-dependent Raman spectra of the BaZrS$_\text{3}$ single crystal, recorded in both parallel and perpendicular configurations for polarization angles ($\theta$) ranging from 0$^\circ$ to 180$^\circ$. The measurements were performed using 785 nm excitation, which provided the highest spectral quality and the most representative response among the available laser wavelengths. A detailed analysis of the polarization-dependent spectra, based on the deconvolution methodology described earlier, enabled the resolution of 12 distinct Raman peaks. The intensity evolution of these peaks follows the expected behavior outlined in Table \ref{tbl:TableWithEquations}, allowing us to assign each peak to either an A$_\text{g}$ or B$_\text{2g}$ mode.

We highlight three representative Raman peaks centered at 51, 90, and 61 cm$^\text{-1}$ in Figures \ref{fig:Figure4}(b), (c), and (d), which exhibit characteristic polarization-dependent intensity trends. The peak at 51 cm$^\text{-1}$ shows periodic intensity variations with a phase shift of $\pi$ (180$^\circ$) in the parallel polarization configuration and $\pi$/2 (90$^\circ$) in the perpendicular configuration, which is typical of A$_\text{g}$ modes. Interestingly, certain A$_\text{g}$ modes -- such as the peaks at 90 cm$^\text{-1}$ (Figure \ref{fig:Figure4}(c)) and 136 cm$^\text{-1}$ -- exhibit a $\pi$/2 (90$^\circ$) phase shift in their intensity profile in the parallel configuration. Consequently, instead of reaching a maximum intensity at the expected polarization angles of -90$^\circ$ and 90$^\circ$, these modes show a maxima at 0$^\circ$. No such shift is observed in the perpendicular polarization configuration, where all A$_\text{g}$ modes behave as expected. This deviation in the polarization-dependent Raman intensity suggests strong electron-phonon coupling in BaZrS$_\text{3}$, aligning with our analysis of the calculated phonon gaps and the vibrational properties reported by Ye et al. \cite{ye2024a}. Such coupling can anisotropically modify the Raman tensor elements, leading to an intensity redistribution and a shift in the expected polarization maxima.

Additionally, the Raman peak at 61 cm$^\text{-1}$ exhibits a sinusoidal intensity variation in the parallel configuration and a cosinusoidal intensity variation in the perpendicular configuration, both with a phase shift of $\pi$/2 (90$^\circ$). This behavior is consistent with the expected behavior of B$_\text{1g}$ modes.

Finally, we use these results from the polarization measurements to assign the symmetry of the Raman peaks, as summarized in Table \ref{table:PeaksAndSymmetries}.

\begin{table*}[ht!]
\small
\caption{Frequency (in cm$^{-1}$) of peaks from Lorentzian fitting of BaZrS$_\text{3}$ Raman spectra and assigned mode symmetry from polarization-dependent Raman measurements compared with theoretical predictions and references.}
\vspace{0.25cm}
\label{table:PeaksAndSymmetries}
\begin{tabular*}{0.8\textwidth}{@{\extracolsep{\fill}}cccccc}
    \multicolumn{5}{c}{this work} & ref \cite{pandey2020a} \\ 
    \cline{1-5} \cline{6-6} \vspace{-0.25cm}\\
    \multicolumn{3}{c}{experimental} & \multicolumn{2}{c}{DFT} \\ 
    \cline{1-3} \cline{4-5} \vspace{-0.25cm}\\
    $\nu_\mathrm{exp}$ (cm$^{-1}$) \vspace{0.1cm} & symmetry & determined using & $\nu_\mathrm{theory}$ (cm$^{-1}$) & symmetry & $\nu_\mathrm{exp}$ (cm$^{-1}$) \\
    51 & A$_\mathrm{g}$ & polarization & 57 & A$_\mathrm{g}$ & 56.5 \\
    61 & B$_\mathrm{2g}$ & polarization & 62 & B$_\mathrm{2g}$ & 64 \\
    74 & A$_\mathrm{g}$ & polarization & 72 & A$_\mathrm{g}$ & 78.6 \\
    72 & B$_\mathrm{3g}$ & 488nm, 638nm & 73 & B$_\mathrm{3g}$ & \\
    77 & B$_\mathrm{2g}$ & polarization & 78 & B$_\mathrm{2g}$ & \\
    85 & B$_\mathrm{1g}$ & 532nm & 83 & B$_\mathrm{1g}$ & 86.7 \\
    88 & B$_\mathrm{2g}$ & polarization & 91 & B$_\mathrm{2g}$ & \\
    90 & A$_\mathrm{g}$ & polarization & 95 & A$_\mathrm{g}$ & 96.5 \\
    111 & B$_\mathrm{1g}$ & 488nm, 532nm & 105 & B$_\mathrm{1g}$ & \\
    136 & A$_\mathrm{g}$ & polarization & 137 & A$_\mathrm{g}$ & 143.1 \\
    145 & B$_\mathrm{3g}$ & 488nm, 532nm & 157 & B$_\mathrm{3g}$ & 154 \\
    160 & A$_\mathrm{g}$ & polarization & 161 & A$_\mathrm{g}$ & 168.3 \\
    176 & B$_\mathrm{2g}$ & 532nm, 638nm & 165 & B$_\mathrm{2g}$ & \\
    182 & B$_\mathrm{1g}$ & 488nm, 532nm & 167 & B$_\mathrm{1g}$ & 192 \\
    190 & B$_\mathrm{2g}$ & polarization & 200 & B$_\mathrm{2g}$ & 208.5 \\
    207 & B$_\mathrm{3g}$ & 488nm & 200 & B$_\mathrm{3g}$ & 208.5 \\
    215 & A$_\mathrm{g}$ & polarization & 205 & A$_\mathrm{g}$ & 217.2 \\
    222 & B$_\mathrm{2g}$ & polarization & 213 & B$_\mathrm{2g}$ & 225 \\
    \multirow{2}{*}{248} & \multirow{2}{*}{B$_\mathrm{1g}$/B$_\mathrm{3g}$} & \multirow{2}{*}{488nm} & 269 & B$_\mathrm{1g}$ & 285 \\
    & & & 270 & B$_\mathrm{3g}$ & 285 \\
    287 & A$_\mathrm{g}$ & 488nm, 532nm & 283 & A$_\mathrm{g}$ & \\
    362 & B$_\mathrm{1g}$ & 488nm, 532nm & 374 & B$_\mathrm{1g}$ & 428 \\
    375 & B$_\mathrm{2g}$ & polarization & 386 & B$_\mathrm{2g}$ & 428 \\
    420 \vspace{0.1cm} & B$_\mathrm{3g}$ & 488nm, 532nm & 405 & B$_\mathrm{3g}$ & \\
    \hline 
\end{tabular*}
\end{table*}

\section{Discussion}

Our results on single-crystal BaZrS$_\text{3}$ provide valuable insights into its optoelectronic potential. Compared to prior reports of broad and multi-component PL emissions, our study demonstrates a strong, band-to-band-dominated PL with a single narrow emission peak. This suggests that high-quality synthesis techniques minimize defect states that would otherwise broaden the PL emission. These findings support the hypothesis that previously reported discrepancies in PL spectra stem from defects and impurities, rather than intrinsic material limitations.

One of the central challenges with BaZrS$_\text{3}$ has been its short carrier lifetime. Our TRPL measurements reveal a characteristic decay time of approximately 1 ns, comparable to other high-quality samples but still limited, likely due to phonon interactions. The presence of phonon gaps in the material, as discussed in the Raman and DFT calculations, indicates that energy transfer from excited carriers to the lattice is enhanced, facilitating non-radiative recombination. This observation aligns with previous work by Li et al. \cite{li2022a}, which showed that polaron formation in BaZrS$_\text{3}$ can enhance non-radiative recombination pathways, leading to additional decay channels that may reduce the lifetime of the photogenerated carriers. This is further corroborated by our polarization Raman measurements, which showed unusual shifts in the intensity of Raman modes with polarization angle, indicating the presence of strong electron-phonon coupling. 

Additionally, the low thermal conductivity characteristic of BaZrS$_\text{3}$, which has been attributed to short phonon lifetimes \cite{agyemang2020a}, suggests that thermal energy cannot efficiently dissipate, leading to localized heating effects that influence carrier recombination rates. Higher local temperatures could elevate phonon-assisted non-radiative recombination pathways, as reported in studies on electron-phonon interactions \cite{alatawi2024a}. Our study further supports this, as the presence of phonon gaps observed in DFT calculations indicates ineffecient heat dissipation mechanisms. While this property may be advantageous for thermoelectric applications, it presents a challenge for optoelectronic applications, where rapid heat dissipation is critical to preventing performance degradation.

Overall, our study provides a more refined understanding of the optoelectronic and vibrational properties of BaZrS$_\text{3}$, particularly in the context of high-quality single-crystal synthesis. The observed strong and narrow PL suggests that synthesis improvements can yield higher-performance materials with fewer non-radiative losses. However, the persistent challenge of short carrier lifetimes due to strong electron-phonon coupling remains a key limitation. Future efforts should focus on strategies to decouple electronic transport from phonon interactions, such as compositional tuning or alloying, to optimize BaZrS$_\text{3}$ for practical optoelectronic applications.

\section{Conclusion}

In summary, we present a comprehensive investigation of the optoelectronic and vibrational properties of high-quality single-crystal BaZrS$_\text{3}$, establishing its potential for optoelectronic applications. Our results confirm a strong band-to-band-dominated photoluminescence (PL) with a single narrow emission peak, suggesting that previous reports of weak or inconsistent PL were influenced by synthesis-dependent defects. Time-resolved PL measurements indicate a carrier lifetime of approximately 1 ns, highlighting the impact of strong electron-phonon coupling on non-radiative recombination. Additionally, Raman spectroscopy and DFT calculations reveal the presence of phonon gaps, further corroborating the role of phonon-assisted carrier decay. Finally, using multiwavelength excitation and polarization-dependent Raman spectroscopy, we present a detailed identification of all Raman-active modes, providing a reliable reference for future studies. 

Despite these promising optoelectronic properties, our study also underscores challenges such as short carrier lifetimes and nontrivial electron-phonon interactions, which limit its device performance potential. Future research should explore compositional tuning and defect passivation strategies to mitigate non-radiative recombination losses and enhance the viability of BaZrS$_\text{3}$ for practical applications. By establishing the intrinsic optoelectronic and vibrational properties of BaZrS$_\text{3}$, this work paves the way for further advancements in chalcogenide perovskites for energy and optoelectronic technologies. 

\section*{Conflicts of interest}
There are no conflicts of interest to declare.

\section*{Acknowledgements}
The work presented here is supported by the Carlsberg Foundation, grant CF24-0200. L.B. and N.M. gratefully acknowledge support from the Deutsche Forschungsgemeinschaft (DFG) under Germany’s Excellence Strategy (EXC 2077, No. 390741603, University Allowance, University of Bremen) and Lucio Colombi Ciacchi, the host of the “U Bremen Excellence Chair Program”. A.O-G, C.P, and N.M. acknowledge support from the NCCR MARVEL, a National Centre of Competence in Research, funded by the Swiss National Science Foundation (Grant number 205602). DFT calculations were performed at the Swiss National Supercomputing Centre (CSCS) under project ID s1295.

\section*{Data availability}
The data that support the findings of this study are available from the corresponding authors upon request. The crystallographic information file (CIF) has been deposited with the CCDC and assigned CSD Number 2465609.

\vfill


\nocite{*}

\bibliography{references}

\end{document}



\title{\Large{SUPPORTING INFORMATION} \\ \vspace{1cm} \large BaZrS$_\text{3}$ Lights Up: The Interplay of Electrons, Photons, and Phonons in Strongly Luminescent Single Crystals}

\author{Rasmus Svejstrup Nielsen}
\email[]{Electronic mail: rasmus.nielsen@empa.ch}
\affiliation{Nanomaterials Spectroscopy and Imaging, Transport at Nanoscale Interfaces Laboratory, Swiss Federal Laboratories for Material Science and Technology (EMPA), Ueberlandstrasse 129, 8600 Duebendorf, Switzerland}

\author{Ángel Labordet Álvarez}
\affiliation{Nanomaterials Spectroscopy and Imaging, Transport at Nanoscale Interfaces Laboratory, Swiss Federal Laboratories for Material Science and Technology (EMPA), Ueberlandstrasse 129, 8600 Duebendorf, Switzerland}
\affiliation{Department of Physics, University of Basel, 4056 Basel, Switzerland}
\affiliation{Swiss Nanoscience Institute, University of Basel, 4056 Basel, Switzerland}

\author{Yvonne Tomm}
\affiliation{Department of Structure and Dynamics of Energy Materials, Helmholtz-Zentrum Berlin für Materialien und Energie, Hahn-Meitner-Platz 1, 14109 Berlin, Germany}

\author{Galina Gurieva}
\affiliation{Department of Structure and Dynamics of Energy Materials, Helmholtz-Zentrum Berlin für Materialien und Energie, Hahn-Meitner-Platz 1, 14109 Berlin, Germany}

\author{Andres Ortega-Guerrero}
\affiliation{Nanotech@surfaces Laboratory, Swiss Federal Laboratories for Material Science and Technology (EMPA), Ueberlandstrasse 129, 8600 Duebendorf, Switzerland}

\author{Joachim Breternitz}
\affiliation{Department of Structure and Dynamics of Energy Materials, Helmholtz-Zentrum Berlin für Materialien und Energie, Hahn-Meitner-Platz 1, 14109 Berlin, Germany}
\affiliation{FH Münster, Department of Chemical Engineering, Stegerwaldstr. 39, 48565 Steinfurt, Germany}

\author{Lorenzo Bastonero}
\affiliation{U Bremen Excellence Chair, Bremen Center for Computational Materials Science, and MAPEX Center for Materials and Processes, University of Bremen, D-28359 Bremen, Germany}

\author{Nicola Marzari}
\affiliation{U Bremen Excellence Chair, Bremen Center for Computational Materials Science, and MAPEX Center for Materials and Processes, University of Bremen, D-28359 Bremen, Germany}
\affiliation{PSI Center for Scientific Computing,
Theory, and Data, and National Centre for Computational Design and Discovery of Novel Materials (MARVEL), 5232 Villigen PSI, Switzerland}
\affiliation{Theory and Simulation of Materials (THEOS), and National Centre for Computational Design and Discovery of Novel Materials (MARVEL), \'Ecole Polytechnique F\'ed\'erale de Lausanne (EPFL), CH-1015 Lausanne, Switzerland}

\author{Carlo Pignedoli}
\affiliation{Nanotech@surfaces Laboratory, Swiss Federal Laboratories for Material Science and Technology (EMPA), Ueberlandstrasse 129, 8600 Duebendorf, Switzerland}

\author{Susan Schorr}
\affiliation{Department of Structure and Dynamics of Energy Materials, Helmholtz-Zentrum Berlin für Materialien und Energie, Hahn-Meitner-Platz 1, 14109 Berlin, Germany}
\affiliation{Institute of Geological Sciences, Freie Universität Berlin, Malteserstr. 74–100, 12249 Berlin, Germany}

\author{Mirjana Dimitrievska}
\email[]{Electronic mail: mirjana.dimitrievska@empa.ch}
\affiliation{Nanomaterials Spectroscopy and Imaging, Transport at Nanoscale Interfaces Laboratory, Swiss Federal Laboratories for Material Science and Technology (EMPA), Ueberlandstrasse 129, 8600 Duebendorf, Switzerland}

\maketitle

\clearpage

\begin{figure*}[t!]
    \centering
    \includegraphics[width=\textwidth,trim={0 0 0 0},clip]{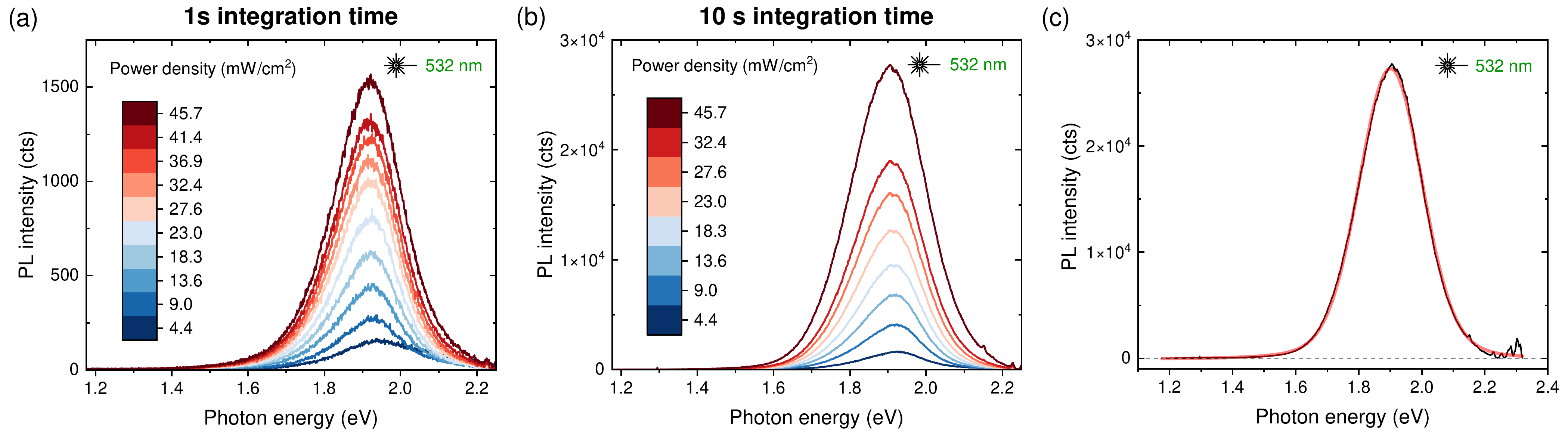}
    \caption{Excitation power-dependent photoluminescence spectra using a 532 nm laser. (a) Data acquired with a 1-second integration time and (b) with a 10-second integration time. (c) Gaussian peak fit of the single-component emission peak.}
    \label{fig:ESI_Figure1}
\end{figure*}

\clearpage

\begin{figure*}[t!]
    \centering
    \includegraphics[width=0.66\textwidth,trim={0 0 0 0},clip]{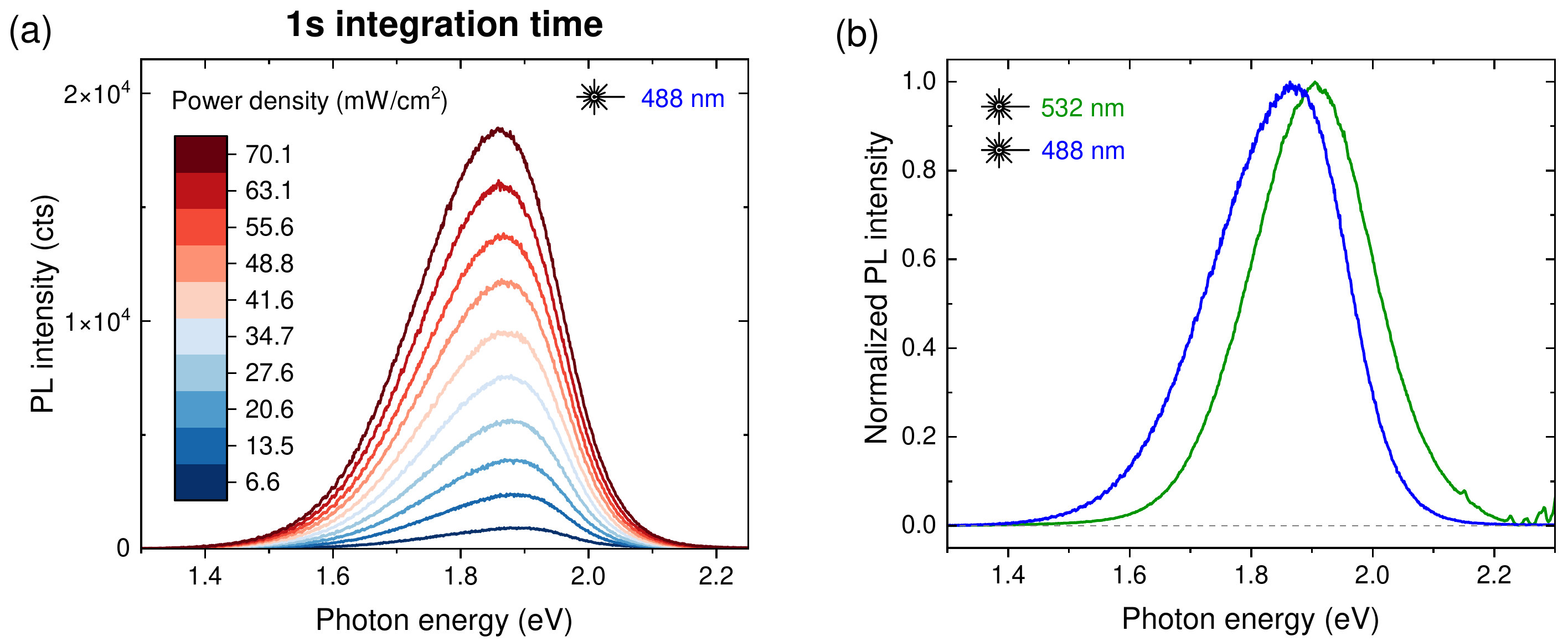}
    \caption{(a) Excitation power-dependent photoluminescence spectra measured with a 488 nm laser. (b) Comparison of photoluminescence spectra measured using 488 nm and 532 nm excitation lasers. A 30 meV redshift in the emission peak is observed, which we attribute to local heating effects and enhanced surface-related recombination under 488 nm excitation.}
    \label{fig:ESI_Figure2}
\end{figure*}

\clearpage

\begin{figure*}[t!]
    \centering
    \includegraphics[width=0.7\textwidth,trim={0 0 0 0},clip]{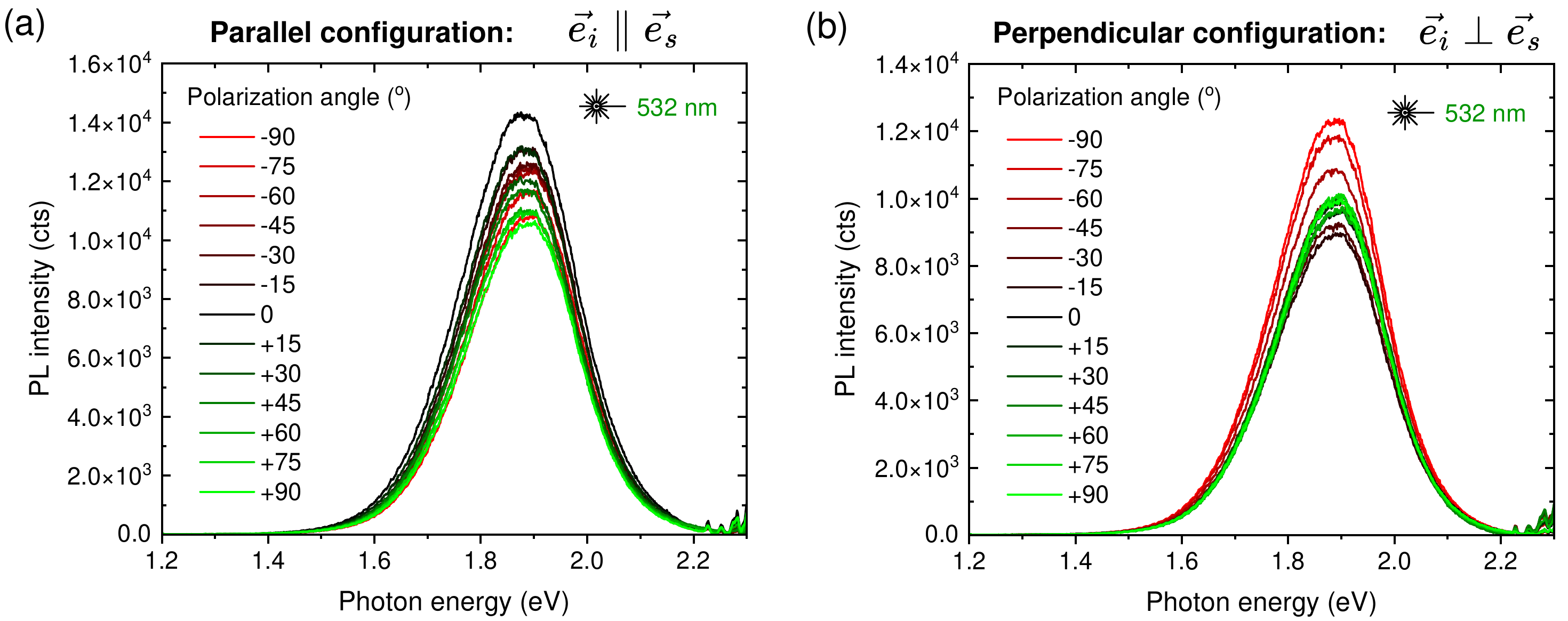}
    \caption{Polarization-dependent photoluminescence spectra measured using a 532 nm excitation laser in (a) parallel and (b) perpendicular configurations.}
    \label{fig:ESI_Figure3}
\end{figure*}

\clearpage

\begin{figure*}[t!]
    \centering
    \includegraphics[width=0.5\textwidth,trim={0 0 0 0},clip]{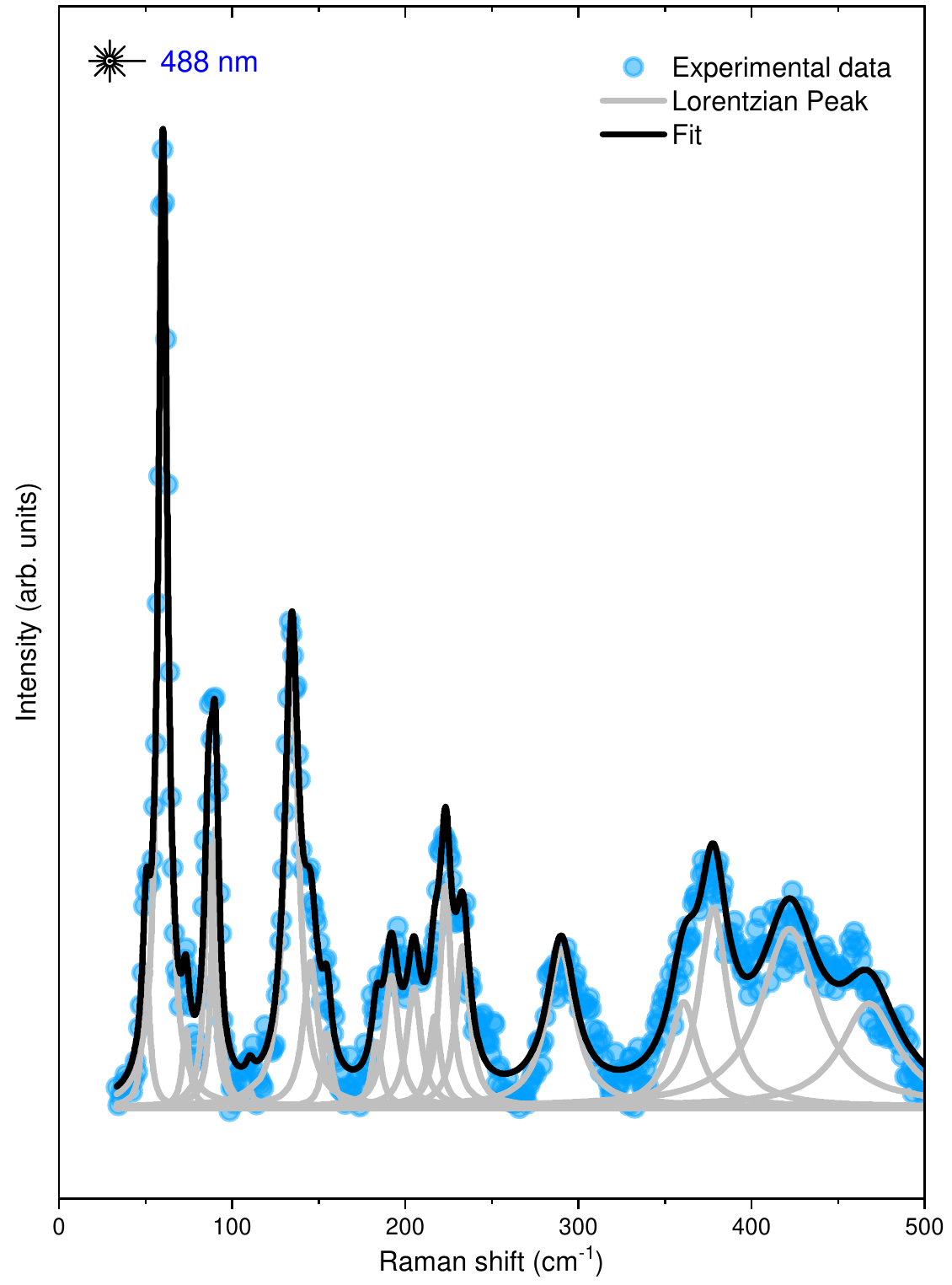}
    \caption{Deconvolution of the Raman spectrum using Lorentzian peak fitting. The spectrum is measured with a 488 nm excitation laser at room temperature.}
    \label{fig:ESI_Figure4}
\end{figure*}

\clearpage

\begin{figure*}[ht!]
    \centering
    \includegraphics[width=0.75\textwidth,trim={0 0 0 0},clip]{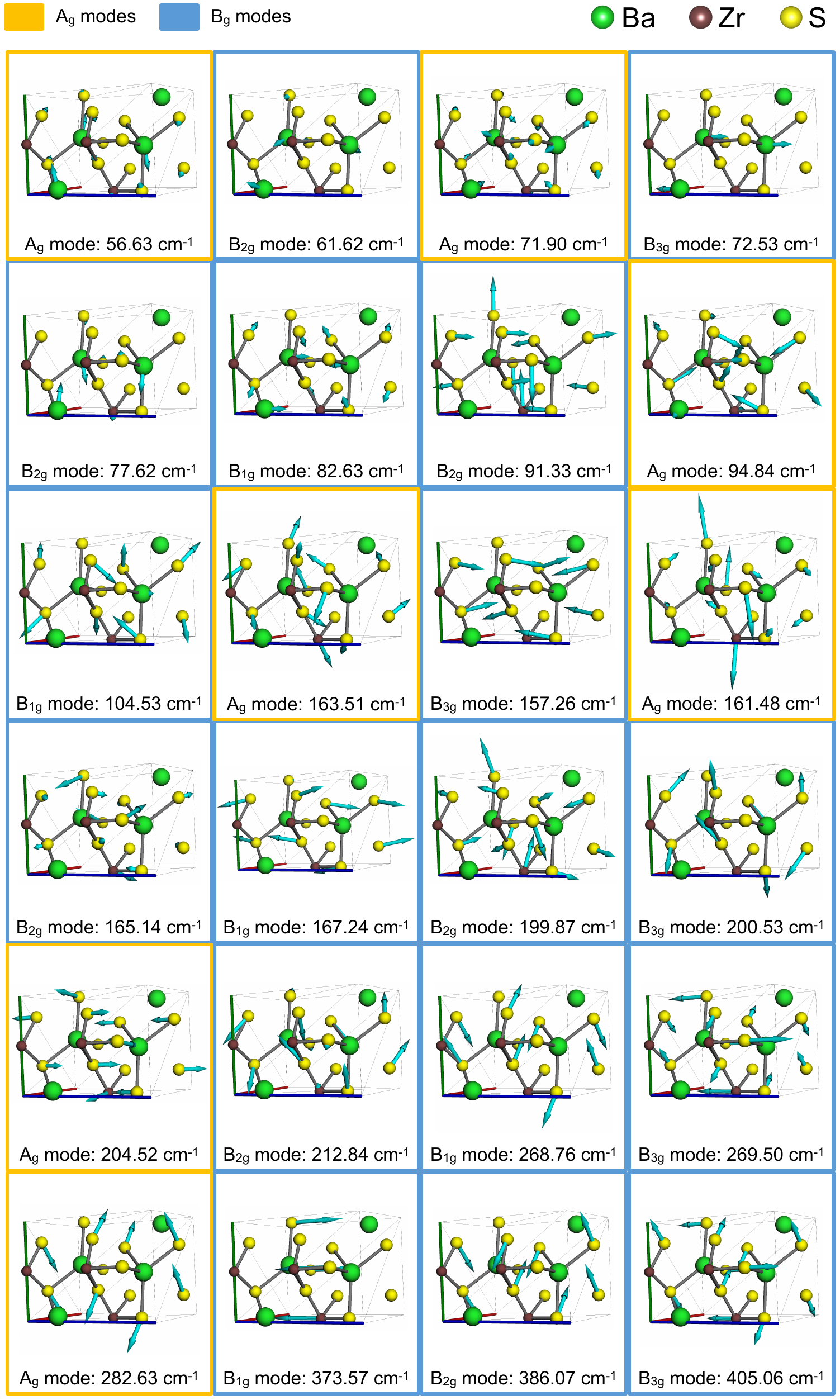}
    \caption{Computed phonon displacement patterns for Raman-active modes of BaZrS$_\text{3}$. The mode symmetries and frequencies (in cm$^\text{-1}$) are listed under each illustration.}
    \label{fig:ESI_Figure5}
\end{figure*}

